\documentclass[a4paper,english,aps]{revtex4}
\usepackage[T1]{fontenc}
\usepackage[latin9]{inputenc}
\usepackage{refstyle}
\usepackage{amsmath}

\makeatletter

\AtBeginDocument{\providecommand\tabref[1]{\ref{tab:#1}}}
\pdfpageheight\paperheight
\pdfpagewidth\paperwidth

\providecommand{\tabularnewline}{\\}
\RS@ifundefined{subsecref}
  {\newref{subsec}{name = \RSsectxt}}
  {}
\RS@ifundefined{thmref}
  {\def\RSthmtxt{theorem~}\newref{thm}{name = \RSthmtxt}}
  {}
\RS@ifundefined{lemref}
  {\def\RSlemtxt{lemma~}\newref{lem}{name = \RSlemtxt}}
  {}

\@ifundefined{textcolor}{}
{%
 \definecolor{BLACK}{gray}{0}
 \definecolor{WHITE}{gray}{1}
 \definecolor{RED}{rgb}{1,0,0}
 \definecolor{GREEN}{rgb}{0,1,0}
 \definecolor{BLUE}{rgb}{0,0,1}
 \definecolor{CYAN}{cmyk}{1,0,0,0}
 \definecolor{MAGENTA}{cmyk}{0,1,0,0}
 \definecolor{YELLOW}{cmyk}{0,0,1,0}
}

\makeatother

\usepackage{babel}
\begin{document}
\begin{abstract}
Certain geometric approximations such as the widely used traditional
shallow-atmosphere, spherical-geoid (TSA-SG) and the deep-atmosphere,
spherical-geoid (DA-SG) approximations boil down to the specification
of a spatial metric tensor. In order to eliminate the leading-order
errors due to the SG and TSA approximations, a sequence of three metric
geometric approximations of increasing accuracy at high altitudes
is obtained. 

Their metric tensors possess a simple, closed-form analytical expression.
The approximations capture to leading order the oblateness of the
planet, the widening of atmospheric columns with height, the horizontal
and vertical variations of gravity and the non-traditional part of
the Coriolis force. Furthermore, for the first two approximations,
the horizontal metric is conformal (proportional) to the spherical
metric, which simplifies analytical and numerical formulations of
the equations of motion. 
\end{abstract}

\title{Simple geometric approximations for global atmospheres on moderately
oblate planets}

\author{Thomas Dubos}

\affiliation{IPSL/Laboratoire de Météorologie Dynamique, École Polytechnique,
Palaiseau, France}
\maketitle

\section{Introduction}

Due to centrifugal effects, rotating planets are oblate rather than
spherical. More generally, geoids (isosurfaces of geopotential $\Phi$)
are quite accurately approximated as oblate ellipsoids. Departure
from sphericity is measured by the flattening $\varepsilon$ (also
called ellipticity or oblateness) of a reference ellipsoid which approximates
either a geoid close to the solid surface of a rocky planet or a constant
pressure surface, typically at $1000\,hPa$. Flattening $\varepsilon$,
defined as the relative difference between the semi-major and semi-minor
axes $a$ and $b=a(1-\varepsilon)$ of the reference ellipsoid, reaches
about $1/10$ for giant planets of the Solar system \citep{Helled2013INTERIOR},
but is only about $\varepsilon\simeq1/300\ll1$ for Earth. In Cartesian
coordinates $(x,y,z)$ centered on the planet's center of mass, with
the $z$ axis aligned with the axis of rotation of the planet, an
expression accurate to $O(\varepsilon)$ of the geopotential is \citep{White2008Spheroidal}
: 
\begin{align}
\Phi=\Phi_{0}\frac{a}{r} & \left[1-\left(\varepsilon-\frac{m}{2}\right)\left(\frac{r}{a}\right)^{-2}\left(\sin^{2}\chi-\frac{1}{3}\right)\right.\label{eq:geopot}\\
 & \left.+\frac{m}{2}\left(\frac{r}{a}\right)^{3}\cos^{2}\chi\right]\nonumber 
\end{align}
where $\lambda,\,\chi,\,r$ are spherical coordinates such that $(x,\,y,\,z)=(r\cos\lambda\cos\chi,\,r\sin\lambda\cos\chi,\,r\sin\chi)$,
\[
\Phi_{0}=\frac{\Gamma M_{E}}{a}
\]
with $\Gamma$ Newton's gravitational constant, $M_{E}$ the mass
of the Earth or planet and 
\[
m=\frac{\Omega^{2}a^{2}}{\Phi_{0}}=O(\varepsilon).
\]
The ratio $m/\varepsilon$ is $O(1)$ and its precise value depends
on the repartition of mass inside the planet (for instance $m/\varepsilon=4/5$
for a uniform-density planet). 

Because $\varepsilon\ll1$, it is quite accurate to neglect the non-sphericity
of geoids when formulating the equations of atmospheric or oceanic
motion for theoretical or numerical purposes. This omission of $O(\varepsilon)$
terms in the equations of motion defines the spherical-geoid (SG)
approximation. If no further approximations are made, the resulting
equations are the so-called deep-atmosphere (DA) equations of motion
\citep{White1995Dynamically}. Assuming spherical geoids simplifies
the expression of metric coefficients involved in the equations of
motion and suppresses the horizontal variations of the acceleration
of gravity. Another parameter whose smallness is often used to further
simplify equations of motion is the relative thickness of the atmosphere
$\tau=H/a=c^{2}/\Phi_{0}$ where $c$ is a typical speed of sound
and $H=c^{2}/(\Phi_{0}/a)$ is the height scale. On Earth $\tau\simeq1/500=O(\varepsilon)$.
Loosely speaking, neglecting terms of order $O(\tau)$ in the equations
of motion defines the traditional shallow-atmosphere (TSA) approximation
(see \citet{Tort2014Dynamically} for a more in-depth discussion).

Although the SG and TSA approximations are quite accurate, it may
be desirable to avoid the errors they incur. \citet{Benard2015Assessment}
has evaluated global forecast errors due to the SG approximation using
idealized shallow-water numerical experiments. An important conclusion
is that these errors could be comparable in magnitude to those due
to the shallow-atmosphere approximation. Indeed, TSA errors are formally
$O(\tau)$ while SG errors are $O(\varepsilon)$, and $\tau\sim\varepsilon$.
Therefore, solving DA-SG equations of motion may not be really more
accurate that solving the TSA-SG equations until the SG approximation
is also relaxed.\\

A handful of operational numerical models of the global atmosphere
do relax the TSA approximation and solve the deep-atmosphere equations
\citep{Wood2014Inherently,Tort2015Energyconserving}. But to the best
of the author's knowledge, a spherical geoid and horizontally-uniform
gravity are assumed in all current operational atmospheric models.
One reason is of course that SG errors are in principle small. In
fact they might not be that small on some planets, especially gaseous
giants. Conversely on Earth, where these errors are presumably very
small indeed, the expected quality of numerical modelling is also
very high. Another compelling reason to retain the SG approximation
is that, to date, no system of three-dimensional equations of motion
has been proposed that would include the effects of oblateness in
a fully satisfactory way and would also be straightforward to implement
in existing models. \\

Previous efforts to relax the SG approximation tackle two aspects
of the problem. One aspect is the construction of orthogonal geopotential
coordinates, that is of a mapping $(\lambda,\phi,\Phi)\mapsto\mathbf{r}=(x,y,z)$
where $\lambda$ is longitude and $\phi$ is a latitudinal coordinate.
This mapping should be orthogonal (or at least such that $\partial\mathbf{r}/\partial\Phi$
be orthogonal to $\partial\mathbf{r}/\partial\phi$ and $\partial\mathbf{r}/\partial\lambda$)
in order to avoid the pollution of the horizontal momentum budget
by large gravitational terms that would be compensated by a similarly
large pressure gradient, obscuring the interesting dynamics and potentially
causing numerical inaccuracies. Even given the relatively simple,
closed-form expression (\ref{eq:geopot}), constructing orthogonal
geopotential coordinates is a non-trivial problem. Previous attempts
rely on geometrical constructions of families of ellipsoids \citep{Benard2014Oblatespheroid,Staniforth2015Geophysically}
but the actual computation of the mapping involve expansions in series
or some sort of iterative procedure. Recently, \citet{Tort2014Usual}
(hereafter TD14) devised a perturbative procedure to construct a mapping
$\mathbf{r}(\lambda,\phi,\Phi)$ that is only quasi-orthogonal, i.e.
$\partial\mathbf{r}/\partial\Phi\cdot\partial\mathbf{r}/\partial\lambda=0$
but $\partial\mathbf{r}/\partial\Phi\cdot\partial\mathbf{r}/\partial\phi=O(\varepsilon^{n})$
for $n$ as large as desired. They obtain a simple, closed-form expression
for $n=2$.

The second aspect of the problem is to formulate the equations of
motion in coordinates $(\lambda,\phi,\Phi)$. \citet{Gates2004Derivation}
starts from the equations of motion in Cartesian coordinates. He then
relates the orthonormal basis associated to coordinates $(x,y,z)$
to the local orthonormal basis associated to $(\lambda,\phi,\Phi)$,
obtains formulae for the spatial derivatives of the latter, and finally
substitutes into the equations of fluid motion in Cartesian coordinates.
This procedure is not only very tedious, it also misses a fundamental
point : ultimately, it is not so much the mapping $\mathbf{r}(\lambda,\phi,\Phi)$
that matters. In fact, as shown by \citet{White2012Consistent} for
orthogonal geopotential coordinates and by TD14 for general non-orthogonal
coordinates, knowing only the metric tensor $g_{ij}=\partial_{i}\mathbf{r}\cdot\partial_{j}\mathbf{r}$
(where $i,j=\lambda,\phi,\Phi$) and the covariant components of planetary
velocity $R_{i}=(\Omega\mathbf{e}_{z}\times\mathbf{r})\cdot\partial_{i}\mathbf{r}$
(where $i=\lambda,\phi,\Phi$) is sufficient to formulate the equations
of motion in coordinates $(\lambda,\phi,\Phi)$. All that needs to
be done is to derive explicitly $g_{ij}$ and $R_{i}$ and inject
them into the TD14 generic form of the equations of motion. 

Because the TD14 mapping is not strictly orthogonal, the resulting
equations of motion would have $O(\varepsilon^{n})$ gravitational
terms in the horizontal momentum budget. If one is willing to accept
errors of order $O(\varepsilon^{n})$ (instead of the $O(\varepsilon)$
errors involved in the SG approximation), these terms can be neglected.
More precisely, following the procedure advocated by TD14 for the
sake of dynamical consistency (in the sense defined by \citealp{White1995Dynamically}),
$g_{ij}$ and $R_{i}$ can be approximated to accuracy $O(\varepsilon^{n-1})$.
By neglecting more generally all $O(\varepsilon^{n})$ terms in $g_{ij}$
and $R_{i}$, the expression of $g_{ij}$ is further simplified. \\

The main purpose of this work is to provide a practical way to incorporate
the effects of planetary oblateness into the equations of atmospheric
and oceanic motion. To this end, the simplest possible expressions
of $g_{ij}$ and $R_{i}$ that take into account planetary oblateness
to leading order are sought. Following the procedure sketched above,
expressions of $g_{ij}$ accurate to $O(\varepsilon)$ are obtained
in Appendix C. In section 2, further simplifications to these expressions
are devised. These simplified expressions are still accurate to $O(\varepsilon)$
at altitudes $O(\varepsilon a)$, where most of the atmospheric mass
resides if $\tau\le O(\varepsilon)$ , but less accurate at altitudes
$O(a)$. Section 3 discusses practical issues, including values of
defining parameters for Earth. A brief section 4 concludes.

\section{Simple geometric approximations}

\subsection{Procedure}

Notice that (\ref{eq:geopot}) uses the same sign convention as \citet{White2008Spheroidal},
while the opposite convention is most often used. It is convenient
and more conventional to define a geopotential $\xi$ that increases
with height and vanishes on the reference ellipsoid : 
\[
\xi=\Phi_{a}-\Phi
\]
where $\Phi_{a}$ is the value of $\Phi$ on the reference ellipsoid.
For the sake of conciseness, units such that $a=1$ and $\text{\ensuremath{\Phi}}_{0}=1$
are adopted in this section and in the appendices. Standard units
are restored at the end of this section. 

\citet{Tort2014Usual} define nearly-orthogonal coordinates where
the third coordinate $\xi$ is, as above, a function of $\Phi$ only
(see Appendix B). As done in appendix C, it is possible to use their
derivation to obtain the corresponding metric tensor truncated to
$O(\varepsilon)$, which by design is orthogonal, i.e. the squared
length $\text{d}l^{2}$ associated to variations of the coordinates
$\lambda,\phi,\xi$ is of the form :
\begin{equation}
\text{d}l^{2}=h_{\lambda}^{2}\text{d}\lambda^{2}+h_{\phi}^{2}\text{d}\text{\ensuremath{\phi}}^{2}+g^{-2}\text{d}\xi^{2}\label{eq:orthogonal}
\end{equation}
where $g(\phi,\xi)=\left\Vert \nabla\Phi\right\Vert =\left\Vert \nabla\xi\right\Vert $
is the local value of gravity and $\phi$ is a coordinate akin to
latitude, to be specified more precisely below. In order to obtain
the simplest possible expressions of $h_{\lambda},\,h_{\phi},\,g$
that capture effects neglected in the TSA-SG approximation, we distinguish
the low-altitude region $\xi=O(\varepsilon)$ from the high-altitude
region $\xi=O(1)$. Simpler expressions, accurate only for $\xi=O(\varepsilon)$
are found by expanding $h_{\lambda},\,h_{\phi},\,g$ in powers of
$\xi$ then truncating to first order in $\varepsilon$. This expansion
is done formally in Appendix D, but here we follow a shortcut : knowing
that this approximate orthogonal metric exists, we construct it step-by
step by invoking physical arguments. While the resulting expressions
are as simple as one can get to capture the effects of oblateness
with $O(\varepsilon)$ accuracy for $\xi=O(\varepsilon)$, they are
fully inaccurate at high altitudes $\xi=O(1)$. Noticing that the
deep-atmosphere expressions for $h_{\lambda},\,h_{\phi},\,g$ are
accurate to $O(1)$ at such altitudes, the previously obtained expressions
are modified in order to restore $O(1)$ accuracy at $\xi=O(1)$ without
loss of accuracy at $\xi=O(\varepsilon$). 

\subsection{Horizontal metric on the reference ellipsoid}

The value $\Phi_{a}$ of $\Phi$ on the reference ellipsoid is obtained
by letting $(\lambda,\chi,r)=(0,0,1)$ in :
\begin{equation}
\Phi=r^{-1}-\left(\varepsilon-\frac{m}{2}\right)r^{-3}\left(\sin^{2}\chi-\frac{1}{3}\right)+\frac{m}{2}r^{2}\cos^{2}\chi,\label{eq:geopot_adim}
\end{equation}
yielding
\[
\Phi_{a}=1+\frac{\varepsilon+m}{3}.
\]
Letting $\chi=\pi/2$ and $\Phi=\Phi_{a}$ in (\ref{eq:geopot_adim})
yields $r=1-\varepsilon+O(\varepsilon^{2})$, confirming that the
semi-minor axis of the reference ellipsoid is indeed $1-\varepsilon$.
On this ellipsoid, there exist many orthogonal coordinate systems
$(\lambda,\phi)$ which are similar to spherical coordinates in that
$\lambda$ is longitude ($\lambda=cst$ is a great circle) and $\phi$
is akin to latitude ($\phi=cst$ is a circle parallel to the Equator).
Appendix A recalls the definition of several variants of latitude
(reduced latitude , geodetic latitude and conformal latitude) which
coincide on a perfect sphere but differ for finite flattening. Rather
than the commonly used \emph{geodetic latitude}, we choose to define
$\phi$ as the \emph{conformal latitude}. Using this latitudinal coordinate,
the scale factors are, to $O(\varepsilon)$ accuracy : 
\begin{align}
h_{\phi} & =1-\varepsilon\sin^{2}\phi\nonumber \\
h_{\lambda} & =h_{\phi}\cos\phi\label{eq:conformal_ellipsoid}
\end{align}
With (\ref{eq:conformal_ellipsoid}), $\text{d}l^{2}=h_{\lambda}^{2}\text{d}\lambda^{2}+h_{\phi}^{2}\text{d}\phi^{2}$
is said to be conformal with respect to the strictly spherical metric\emph{
$\text{d}\phi^{2}+\cos^{2}\phi\text{d}\text{\ensuremath{\lambda}}^{2}$}
because it differs from the latter only by the multiplicative factor
$h_{\phi}^{2}$. This property simplifies numerical formulations of
differential operators (see 3.3).

\subsection{Gravity on the reference ellipsoid}

We have decided to use the conformal latitude $\phi$ of the reference
ellipsoid as latitudinal coordinate. There is no guarantee that the
TD14 latitudinal coordinate coincides with $\phi$ and it is in fact
demonstrated in Appendix C that they differ slightly. Even so, due
to zonal symmetry, there exists a change of latitudinal coordinate
that, on the reference ellipsoid, maps the TD14 latitude to conformal
latitude. Such a change of coordinate preserves the orthogonal character
of the metric, hence (\ref{eq:orthogonal}) is still valid.

Using (\ref{eq:geopot_adim}) one obtains the values of $g$ at the
Poles and Equator of the reference ellipsoid :
\begin{align*}
g_{P} & =\left.\frac{\partial\Phi}{\partial r}\right|_{\chi=\pi/2,r=1-\varepsilon}=1+m\\
g_{E} & =\left.\frac{\partial\Phi}{\partial r}\right|_{\chi=0,r=1}=1-\frac{3}{2}m+\varepsilon
\end{align*}
In between, $g$ varies with latitude as (see (\ref{eq:gravity_ref-1})
in Appendix C) :
\begin{align}
g(\phi) & =1+m-\left(\frac{5}{2}m-\varepsilon\right)\cos^{2}\phi.\label{eq:gravity_ref}
\end{align}
Notice especially that $g_{P}/g_{E}=1+(5/2)m-\varepsilon+O(\varepsilon^{2})$
as expected \citep{White2008Spheroidal}. 

\subsection{Widening of atmospheric columns with height}

Now consider a horizontal displacement along a geoid $\xi=cst\ne0$.
The squared displacement equals $\text{d}l^{2}=h_{\lambda}^{2}\text{d}\lambda^{2}+h_{\phi}^{2}\text{d}\text{\ensuremath{\phi}}^{2}$
where $h_{\lambda},\,h_{\phi}$ are given at $\xi=0$ by (\ref{eq:conformal_ellipsoid}).
For $\xi=O(\varepsilon)$, 
\begin{align*}
h_{\phi} & =1-\varepsilon\sin^{2}\phi+\left.\frac{\partial h_{\phi}}{\partial\xi}\right|_{\xi=0}\xi+O(\varepsilon^{2})\\
h_{\lambda} & =\left(1-\varepsilon\sin^{2}\phi\right)\cos\lambda+\left.\frac{\partial h_{\lambda}}{\partial\xi}\right|_{\xi=0}\xi+O(\varepsilon^{2}).
\end{align*}
The corrections $(\partial h_{\phi}/\partial\xi)\xi,\,(\partial h_{\lambda}/\partial\xi)\xi$
express the widening of atmospheric columns with height. Their dependence
on $\varepsilon$ can, at this order of accuracy, be neglected. Hence
$\partial h_{\phi}/\partial\xi,\,\partial h_{\lambda}/\partial\xi$
can be obtained under the assumption of spherical geoids. In that
case the distance from the center of the planet is $\Phi^{-1}=\left(1-\xi\right)^{-1}$
, $h_{\lambda}=h_{\phi}\cos\phi$ and $h_{\phi}=(1-\xi)^{-1}=1+\xi+O(\varepsilon^{2}).$
Adding the $O(\xi)$ term to (\ref{eq:conformal_ellipsoid}) yields
:
\begin{align}
h_{\phi} & =1+\xi-\varepsilon\sin^{2}\phi,\label{eq:widening}\\
h_{\lambda} & =h_{\phi}\cos\phi.\nonumber 
\end{align}
Clearly, (\ref{eq:widening}) is inaccurate for $\xi=O(1)$. Indeed
in this region of high altitudes, $h_{\phi}$ should coincide at $O(1)$
with its deep-atmosphere expression $h_{\phi}=\Phi^{-1}=(1-\xi)^{-1}+O(\varepsilon)$
. Hence, the expressions :
\begin{align}
h_{\phi} & =(1-\xi)^{-1}(1-\varepsilon\sin^{2}\phi),\label{eq:widening_deep}\\
h_{\lambda} & =h_{\phi}\cos\phi.\nonumber 
\end{align}
restore $O(1)$ accuracy where $\xi=O(1)$ without loss of accuracy
for $\xi=O(\varepsilon)$, since they coincide with (\ref{eq:widening})
at $O(\varepsilon)$ for $\xi=O(\varepsilon)$. 

\subsection{Vertical variation of gravity}

Similarly, variations of $g$ with height, which are a $O(\varepsilon)$
effect for $\xi=O(\varepsilon)$, can be obtained to $O(\varepsilon)$
accuracy in the purely spherical case. In that case $g=r^{-2}=\Phi^{2}=(1-\xi)^{2}=1-2\xi+O(\varepsilon^{2})$.
Adding the $O(\xi)$ term to (\ref{eq:gravity_ref}) yields :
\begin{equation}
g=1-2\xi+m-\left(\frac{5}{2}m-\varepsilon\right)\cos^{2}\phi\label{eq:gravity_shallow}
\end{equation}
and demanding $O(1)$ accuracy for $\xi=O(1)$ yields :
\begin{align}
g & =(1-\xi)^{2}\left(1+m-\left(\frac{5}{2}m-\varepsilon\right)\cos^{2}\phi\right)\label{eq:gravity_deep}
\end{align}

\subsection{Full expressions}

In summary, (\ref{eq:orthogonal},\ref{eq:widening},\ref{eq:gravity_shallow})
define a spatial metric that is accurate to $O(\varepsilon)$ only
up to $\xi=O(\varepsilon)$. (\ref{eq:orthogonal},\ref{eq:widening_deep},\ref{eq:gravity_deep})
define a metric that is accurate to $O(\varepsilon)$ for $\xi=O(\varepsilon)$
and accurate to $O(1)$ for $\xi=O(1)$. Both metrics are horizontally
conformal to the spherical metric. For $O(\varepsilon)$ accuracy
at $\xi=O(1)$, one should use (\ref{eq:dl2_TD14},\ref{eq:gravity_TD14},\ref{eq:h_beta_Phi},\ref{eq:h_lambda_Phi})
with definitions (\ref{eq:RE},\ref{eq:DR},\ref{eq:X},\ref{eq:gE},\ref{eq:Delta_g})
(Appendix C). This more accurate metric is \emph{not} horizontally
conformal to the spherical metric. Restoring usual units, the full
expressions of $h_{\lambda},$ $h_{\phi}$ and $g$ are :
\begin{itemize}
\item approximation I (least accurate):
\end{itemize}
\begin{align*}
h_{\lambda}= & h_{\phi}\cos\lambda\\
h_{\phi}= & a\left(1+\frac{\xi}{\Phi_{0}}-\varepsilon\sin^{2}\phi\right)\\
g= & \frac{\Phi_{0}}{a}\left[1-\frac{2\xi}{\Phi_{0}}+m-\left(\frac{5}{2}m-\varepsilon\right)\cos^{2}\phi\right]
\end{align*}

\begin{itemize}
\item approximation II (intermediate)
\end{itemize}
\begin{align*}
h_{\lambda}= & h_{\phi}\cos\lambda\\
h_{\phi}= & a\left(1-\frac{\xi}{\Phi_{0}}\right)^{-1}\left(1-\varepsilon\sin^{2}\phi\right)\\
g= & \frac{\Phi_{0}}{a}\left(1-\frac{\xi}{\Phi_{0}}\right)^{2}\left[1+m-\left(\frac{5}{2}m-\varepsilon\right)\cos^{2}\phi\right]
\end{align*}

\begin{itemize}
\item approximation III (most accurate)
\end{itemize}
\begin{align*}
h_{\lambda}= & a\left(R_{E}\left(\Phi\right)-\Delta R\left(\Phi\right)\sin^{2}\phi+\Delta\phi\left(\Phi\right)\sin^{2}\phi\right)\cos\phi,\qquad\\
h_{\phi}= & a\left(R_{E}\left(\Phi\right)-\Delta R\left(\Phi\right)\sin^{2}\phi-\Delta\phi\left(\Phi\right)\cos2\phi\right),\\
g= & \frac{\Phi_{0}}{a}\left(g_{E}\left(\Phi\right)+\Delta g\left(\Phi\right)\sin^{2}\phi\right),
\end{align*}
\begin{align*}
\text{where }\Phi= & 1+\frac{\varepsilon+m}{3}-\frac{\xi}{\Phi_{0}},\\
R_{E}(\Phi)= & \Phi^{-1}+\frac{1}{3}\left(\varepsilon-\frac{m}{2}\right)\Phi+\frac{m}{2}\Phi^{-4},\\
\Delta R(\Phi)= & \left(\varepsilon-\frac{m}{2}\right)\Phi+\frac{m}{2}\Phi^{-4},\\
\Delta\phi(\Phi)= & \frac{5m}{6}-\varepsilon+\left(\varepsilon-\frac{m}{2}\right)\Phi-\frac{m}{3}\Phi^{-4},\\
g_{E}(\Phi) & =\Phi^{2}+\frac{1}{3}\left(\varepsilon-\frac{m}{2}\right)\Phi^{4}+2m\Phi^{-1},\\
\Delta g(\Phi) & =-\left(\varepsilon-\frac{m}{2}\right)\Phi^{4}+2m\Phi^{-1}.
\end{align*}

\section{Practical considerations}

\subsection{Defining parameters and their values}

The shape of the Earth and its gravitational field are known with
a very high accuracy, and geodetic systems have been very precisely
defined, among which the World Geodetic System 1984 (WGS84) is perhaps
the most relevant at global scale \citep{nima:2000}. For terrestrial
applications, it is therefore desirable to specify values of the parameters
$a,\,\Phi_{0},\,\Omega,\dots$ that match as closely as possible those
of WGS84, some of which are listed in \tabref{WGS84-parameters}.
A perfect match is not possible since WGS84 has better than $O(\varepsilon)$
accuracy. 

We suggest to take as fundamental parameters the semi-major and semi-minor
axes $a,b$ , the gravitational parameter $\Gamma M_{E}$ and the
sidereal period $T=2\pi/\Omega$, which are known with good accuracy
for other planets too, including giant planets such as Saturn and
Jupiter, for which the reference ellipsoid approximates the isobaric
surface $p=10^{5}Pa$. Table \ref{tab:Proposed-parameters} presents
possible values for $a,\,b,\,\Gamma M_{E}$ and $T$ for the Earth,
Saturn and Jupiter. The Earth values of $a$ and $b$ have been rounded
to $1m$ accuracy. In addition to $a,\,b,\,\Gamma M_{E},\,T$, the
derived quantities $\text{\ensuremath{\Omega}},\,\varepsilon,\,m,\,g_{E}$
and $g_{p}$ are computed. It is is seen that the terrestrial values
of gravity at the Equator and Poles match very well, although not
perfectly, those of WGS84. Furthermore large horizontal variations
of $g$ occur on Saturn and Jupiter.

Similarly, one has to define precisely the relation between $\phi$
and geodetic latitude $\phi_{g}$. Since geographical databases typically
refer to $\phi_{g}$, conversions between $\phi_{g}$ and $\phi$
are required when processing inputs and outputs of a numerical model.
There exists a closed-form relationship between $\phi_{g}$ and the
conformal latitude, but it is rather involved and not easily invertible.
The approximate relationship $\phi=\phi_{g}-2\varepsilon\cos\phi_{g}\sin\phi_{g}$
could be used as a definition of $\phi$, which should probably be
called pseudo-conformal latitude, since it differs slightly from the
true conformal latitude. To obtain $\phi_{g}$ from $\phi$, one may
iterate $\phi_{g}\leftarrow\phi-2\varepsilon\sin\phi_{g}\cos\phi_{g}$
from an initial value $\phi_{g}\leftarrow\phi$, or use a few Newton
iterations.

\begin{table}
\begin{centering}
\begin{tabular}{|c|c|}
\hline 
$\Omega$ & $7.292115\times10^{-5}rad\cdot s^{-1}$\tabularnewline
\hline 
$a$ & $6378137.0\,m$\tabularnewline
\hline 
$b$ & $6356752.3142\,m$\tabularnewline
\hline 
$\Gamma M_{E}$ & $3.986004418\times10^{14}\,m^{3}s^{-2}$\tabularnewline
\hline 
$g_{P}$ & $9.8321849378\,m\,s^{-2}$\tabularnewline
\hline 
$g_{E}$ & $9.7803253359\,m\,s^{-2}$\tabularnewline
\hline 
\end{tabular}
\par\end{centering}
\caption{Terrestrial value of important physical quantities according to the
WGS84 geodetic system \citep{nima:2000}. \label{tab:WGS84-parameters}}
\end{table}

\begin{table*}
\begin{centering}
\begin{tabular}{|c||c|c|c||c|}
\hline 
 & Earth & Jupiter & Saturn & Unit\tabularnewline
\hline 
\hline 
$a$ & $6378.137$ & $71492$ & $60268$ & $km$\tabularnewline
\hline 
$b$ & $6356.752$ & $66854$ & $54364$ & $km$\tabularnewline
\hline 
$\Gamma M_{E}$ & $3.9860\times10^{14}$ & $12.6687\times10^{16}$ & $3.7931\times10^{16}$ & $m^{3}s^{-2}$\tabularnewline
\hline 
$T$ & $23.93447$ & $9.9250$ & $10.656$ & $h$\tabularnewline
\hline 
\hline 
$\Omega=2\pi/T$ & $7.292115\times10^{-5}$ & $1.7585\times10^{-4}$ & $1.6379\times10^{-4}$ & $s^{-1}$\tabularnewline
\hline 
$\varepsilon=(a-b)/a$ & $0.0033528$ & $0,06487$ & $0,09796$ & \tabularnewline
\hline 
$m=a^{3}\Omega^{2}/(\Gamma M_{E})$ & $0.0034614$ & $0.08919$ & $0.1548$ & \tabularnewline
\hline 
$g_{P}=$$\frac{\Phi_{0}}{a}\left(1+m\right)$ & $9.83219\text{}$ & $27.00$ & $12.06$ & $m\,s^{-2}$\tabularnewline
\hline 
$g_{E}=\frac{\Phi_{0}}{a}\left(1-\frac{3}{2}m+\varepsilon\right)$ & $9.78025$ & $23.08$ & $9.04$ & $m\,s^{-2}$\tabularnewline
\hline 
\end{tabular}
\par\end{centering}
\caption{Proposed parameters for Earth, Jupiter and Saturn. Values of $a,\,b,\,\Gamma M_{E}$
and sidereal period $T$ closely match those of the WGS84 system for
Earth \citep{nima:2000}. For Jupiter they are taken from \protect\url{https://nssdc.gsfc.nasa.gov/planetary/factsheet/jupiterfact.html}
and for Saturn from \protect\url{https://nssdc.gsfc.nasa.gov/planetary/factsheet/saturnfact.html}
(slightly update values of $a$ and $b$ for Saturn are given in \citealp{Helled2013INTERIOR}).
Other quantities $\Omega,\,\varepsilon,\,m,\,g_{P},g_{E}$ are deduced
from $a,\,b,\,\Gamma M_{E},\,T$. While actual implementations should
compute them to machine accuracy, rounded values are presented in
the table.\label{tab:Proposed-parameters}}
\end{table*}

\subsection{Equations of motion}

$\Omega,\,h_{\lambda},\,h_{\phi}$ and $g$ being fully specified,
\citet{White2012Consistent} provide equations (A.10-A.12) which prognose
the ``physical'' velocity components $(u,v,w)=(h_{\lambda}u^{\lambda},\,h_{\phi}u^{\phi},g^{-1}u^{\xi})$
where the contravariant velocity components $(u^{\lambda},\,u^{\phi},u^{\xi})$
are defined as the Lagrangian derivatives of $(\lambda,\phi,\xi)$
. As noted in TD14, these equations are a special case of those derived
in TD14 in non-orthogonal curvilinear coordinates, provided the Jacobian
$J$ (converting between density $\rho$ and pseudo-density $\mu=\rho J$
involved in the flux-form mass budget) is defined as 
\begin{equation}
J=h_{\lambda}h_{\phi}g^{-1}.\label{eq:Jacobian}
\end{equation}
and the covariant components of planetary velocity are defined as
$(R^{\lambda},R^{\phi},R^{\xi})=(\Omega,0,0)$ hence $(R_{i})=(g_{ij}R^{j})=(\Omega h_{\lambda}^{2},0,0)$:
\begin{equation}
R_{\lambda}=\Omega h_{\lambda}^{2}.\label{eq:planetvel}
\end{equation}

Importantly, definition (\ref{eq:planetvel}) implies that $R_{\lambda}$
depends on the vertical coordinate $\xi$. This property restores
the non-traditional part of the Coriolis force, which is neglected
in the TSA-SG approximation, for which $R_{\lambda}=\Omega a^{2}\cos^{2}\phi$
does not depend on the vertical coordinate \citep{Tort2014Dynamically}.

For practical purposes, especially for numerical modeling, it may
be preferable to use other forms than that given in \citet{White2012Consistent},
forms that would prognose covariant or contravariant components, such
as the flux form or curl form found in TD14, and/or use non-Eulerian
coordinates (e.g. \citealp{Dubos2014Equations}). 

\subsection{Benefit of horizontally conformal coordinates}

While approximations I and II are less accurate than approximation
III, their expression is simpler. Especially, $h_{\lambda}=h_{\phi}\cos\phi$.
Practical benefits of this property are discussed here. Consider as
an illustrative example, simpler than the full equations of fluid
motion, the Poisson problem :
\begin{align}
\Delta p & =f\qquad\text{where}\label{eq:Poisson}\\
\Delta & =\frac{1}{J}\left(\frac{\partial}{\partial\lambda}Jh_{\lambda}^{-2}\frac{\partial}{\partial\lambda}+\frac{\partial}{\partial\phi}Jh_{\phi}^{-2}\frac{\partial}{\partial\phi}+\frac{\partial}{\partial\xi}Jg^{2}\frac{\partial}{\partial\xi}\right)\label{eq:Laplace}
\end{align}
The fields $p$ and $f$, rather than functions of $\lambda,\phi,\xi$,
can be regarded as functions of $\mathbf{x}$ and $\xi$ , where $\mathbf{x}$
is the position on the unit sphere corresponding to longitude $\lambda$
and latitude $\phi$. If $h_{\lambda}=h_{\phi}\cos\phi$, (\ref{eq:Laplace})
can be rewritten as :
\begin{align}
\Delta= & \frac{1}{\tilde{J}}\left(\nabla\cdot\left(\tilde{J}h_{\phi}^{-2}\nabla\right)+\frac{\partial}{\partial\xi}\left(\tilde{J}g^{2}\frac{\partial}{\partial\xi}\right)\right),\label{eq:Laplace_conformal}\\
\text{where }\tilde{J}= & g^{-1}h_{\phi}^{2}\nonumber 
\end{align}
and $\nabla\cdot$ and $\text{\ensuremath{\nabla}}$ are the divergence
and gradient operators on the unit sphere. In TSA-SG geometry, $g,\,h_{\phi}=a$
and $\tilde{J}=g^{-1}a^{2}$ are constants while in the DA-SG geometry
they depend on altitude but not on latitude. Hence the main change
introduced by approximations I and II is that metric factors become
latitude-dependent. No additional operator is introduced beyond $\nabla,\,\nabla\cdot,\,\nabla\times$
and the $\perp$ operator that rotates a vector tangent to the unit
sphere by an angle of $\pi/2$. Upgrading a numerical solver for the
TSA-SG or DA-SG equations to geometries I or II should be essentially
a matter of replacing constant or height-dependent metric factors
by similar factors that depend also on latitude. 

\section{Conclusion}

A sequence of three increasingly accurate metric tensors have been
obtained, that capture to leading order the oblateness of the planet,
the widening of atmospheric columns with height, the horizontal and
vertical variations of gravity and the non-traditional part of the
Coriolis force. Given the presumably small magnitude of the errors
due to the SG approximation, these geometric approximations have been
developed with simplicity and ease of numerical implementation in
mind. Hopefully this work will facilitate the development of even
more accurate global atmospheric solvers.

As a final remark, notice that the geometric approximations I-III
could be applied as well to ocean modeling. In this case, since $H/a\sim10^{-4}$
is an order of magnitude smaller than $\varepsilon$, vertical variations
of $h_{\lambda},\,h_{\phi}$ and $g$ may be neglected without much
loss of accuracy. Expressions for $h_{\lambda},\,h_{\phi},\,g$ would
then reduce to (\ref{eq:conformal_ellipsoid},\ref{eq:gravity_ref}).
On the other hand, in order to retain the non-traditional component
of the Coriolis force, one would expand expression (\ref{eq:planetvel})
of $R_{\lambda}$ to first order in $\xi$ as in \citet{Tort2014Dynamically},
yielding $R_{\lambda}=\Omega a^{2}\left(1+2\xi/\Phi_{0}\right)\cos^{2}\lambda$.
Notice that this $R_{\lambda}\neq\Omega h_{\lambda}^{2}$, so that
the equations of motion should be obtained from TD14 rather than \citet{White2012Consistent}. 

\bibliographystyle{wileyqj}
\bibliography{group-13463}

\appendix

\section{Latitudes on an ellipsoid}

A simple set of orthogonal longitude-latitude coordinates $(\lambda,\phi_{p})$
on an ellipsoid of semi-major axis $1$ and semi-minor axis $1-\varepsilon$
is defined by the mapping :

\begin{equation}
(x,y,z)=\left(\cos\lambda\cos\phi_{p},\sin\lambda\cos\phi_{p},(1-\varepsilon)\sin\phi_{p}\right).\label{eq:parametric_latitude}
\end{equation}
Coordinate $\phi_{p}$ is then called the \emph{reduced or parametric
latitude}. The metric in these coordinates is, assuming $\varepsilon\ll1$,
$\text{d}l^{2}=\text{d}l_{parametric}^{2}+O(\varepsilon^{2})$ where
:
\[
\text{d}l_{parametric}^{2}=\cos^{2}\phi_{p}\text{d}\lambda^{2}+\left(1-2\varepsilon\cos^{2}\phi_{p}\right)\text{d}\phi_{p}^{2}.
\]
The \emph{geodetic latitude} $\phi_{g}$ is defined as the angle between
the symmetry axis $(0,0,1)$ and the poleward horizontal vector. Using
(\ref{eq:parametric_latitude}), one finds for $\varepsilon\ll1$
:
\begin{align}
\phi_{p} & =\phi_{g}-\varepsilon\cos\phi_{g}\sin\phi_{g}+O(\varepsilon^{2})\label{eq:geodetic}\\
\phi_{g} & =\phi_{p}+\varepsilon\cos\phi_{p}\sin\phi_{p}+O(\varepsilon^{2}).\nonumber 
\end{align}

It is seen that $\phi_{p}$ and $\phi_{g}$ differ by the small latitude-dependent
angle $\varepsilon\cos\phi_{p}\sin\phi_{p}$. Different flavors of
latitude differ by a similar amount, with different constant prefactors.
Consider especially the latitude $\phi_{c}$ such that
\begin{align}
\phi_{p} & =\phi_{c}+\varepsilon\cos\phi_{c}\sin\phi_{c}.\label{eq:conformal}
\end{align}
Using $\text{d}\phi_{p}=\left(1+\varepsilon\cos2\phi_{c}\right)\text{d}\phi_{c}$,
$\cos\phi_{p}=\left(1-\varepsilon\sin^{2}\phi_{c}\right)\cos\phi_{c}+O(\varepsilon^{2})$
one finds $\text{d}l^{2}=\text{d}l_{conf}^{2}+O(\varepsilon^{2})$
where :
\begin{align*}
\text{d}l_{conf}^{2} & =\left(1-\varepsilon\sin^{2}\phi_{c}\right)^{2}\left(\cos^{2}\phi_{c}\text{d}\lambda^{2}+\text{d}\phi_{c}^{2}\right)
\end{align*}
The metric $\text{d}l_{conf}^{2}$ is said to be conformal to the
spherical metric $\cos^{2}\phi\text{d}\lambda^{2}+\text{d}\phi^{2}$
because they are related by a scaling factor, here $\left(1-\varepsilon\sin^{2}\phi\right)^{2}$.
Accordingly, $\phi_{c}$ coincides, with accuracy $O(\varepsilon)$,
with the so-called \emph{conformal latitude}, and we shall conflate
the two here. Comparing (\ref{eq:geodetic}) and (\ref{eq:conformal}),
one finds that conformal latitude $\phi_{c}$ and geodetic latitude
$\phi_{g}$ are related by :
\begin{align*}
\phi_{c} & =\phi_{g}-2\varepsilon\cos\phi_{g}\sin\phi_{g}+O(\varepsilon^{2})\\
\phi_{g} & =\phi_{c}+2\varepsilon\cos\phi_{c}\sin\phi_{c}+O(\varepsilon^{2}).
\end{align*}

Finally, let us consider a general ellipsoid, of semi-major axis $A$
and semi-minor axis $A-\Delta A$ (with $\varepsilon=\Delta A/A\ll1$),
and a latitudinal coordinate $\phi_{X}$ on this ellipsoid. Scaling
the above expressions by $A$ :
\begin{align}
\text{d}l^{2}= & h_{X}^{2}\text{d}\phi_{X}^{2}+h_{\lambda}^{2}\text{d}\lambda^{2}+O(\varepsilon)^{2}\label{eq:metric_ellipsoid_1}\\
\text{where }h_{X}= & A-\Delta A\,\sin^{2}\phi_{X}-A\Delta\phi\,\cos2\phi_{X}\label{eq:metric_ellipsoid_2}\\
h_{\lambda}= & \left(A-\Delta A\,\sin^{2}\phi_{X}+A\Delta\phi\,\sin^{2}\phi_{X}\right)\cos\phi_{X}\label{eq:metric_ellipsoid_3}\\
\phi_{X}= & \phi_{c}+\Delta\phi\,\cos\phi_{c}\sin\phi_{c},\label{eq:metric_ellipsoid_4}
\end{align}
where $\Delta\phi=O(\varepsilon)$ is a constant that defines the
relationship between $\phi_{X}$ and $\phi_{c}$. Conversely, if the
metric coefficients $h_{X},\,h_{\lambda}$ can be written in the form
(\ref{eq:metric_ellipsoid_1}-\ref{eq:metric_ellipsoid_4}), the parameters
$A,\,\Delta A$ and $\Delta\phi$ are readily identified and coordinate
$\phi_{X}$ can be related to $\phi_{c}$.

\section{TD14 mapping}

We reproduce here, with adapted notations, the TD14 construction of
quasi-orthogonal coordinates $(\lambda,\beta,\Phi)$ with $\lambda$
longitude, $\beta$ a latitude to defined precisely later, and $\Phi$
geopotential. These coordinates map to the position in 3D Cartesian
space :
\begin{align}
\mathbf{r}= & \left(\Phi^{-1}-\varepsilon\frac{\partial\psi}{\partial\Phi}\right)\mathbf{e}_{r}-\varepsilon\Phi^{-1}\frac{\partial\psi}{\partial\beta}\mathbf{e}_{\beta}\label{eq:map_TD15}
\end{align}
\begin{align*}
\text{where }\text{\ensuremath{\mathbf{e}}}_{r}= & \left(\cos\lambda\cos\beta,\,\sin\lambda\cos\beta,\,\sin\beta\right),\\
\text{\ensuremath{\mathbf{e}}}_{\beta}= & \left(-\cos\lambda\sin\beta,\,-\sin\lambda\sin\beta,\,\cos\beta\right).
\end{align*}
where again we assume $a=1$ and $\Phi_{0}=1$, and $\psi$ is to
be determined below. The form adopted in (\ref{eq:map_TD15}) ensures
that $(\partial\text{\ensuremath{\mathbf{r}}/\ensuremath{\partial\beta})}\cdot(\partial\text{\ensuremath{\mathbf{r}}/\ensuremath{\partial\Phi})}=O(\varepsilon^{2})$.
Now (\ref{eq:map_TD15}) implies $\chi-\beta=O(\varepsilon)$ and
\begin{equation}
r=\Phi^{-1}-\varepsilon\frac{\partial\psi}{\partial\Phi}+O(\varepsilon^{2}).\label{eq:r_Psi}
\end{equation}
Using $\chi-\beta=O(\varepsilon)$, (\ref{eq:geopot_adim}) yields
:
\begin{align}
r & =R(\Phi)+O(\varepsilon^{2})\nonumber \\
\text{where}\qquad R(\Phi) & =R_{E}(\Phi)-\sin^{2}\beta\Delta R(\Phi),\label{eq:R_Phi}\\
R_{E}(\Phi) & =\Phi^{-1}+\frac{1}{3}\left(\varepsilon-\frac{m}{2}\right)\Phi+\frac{m}{2}\Phi^{-4},\label{eq:RE}\\
\Delta R(\Phi) & =\left(\varepsilon-\frac{m}{2}\right)\Phi+\frac{m}{2}\Phi^{-4}.\label{eq:DR}
\end{align}
Substituting (\ref{eq:R_Phi}) in (\ref{eq:r_Psi}) then yields $\psi$
hence $\partial\psi/\partial\beta$ : 
\begin{align}
\varepsilon\frac{\partial\psi}{\partial\Phi} & =\left(\varepsilon-\frac{m}{2}\right)\left(\sin^{2}\beta-\frac{1}{3}\right)\Phi-\frac{m}{2}\Phi^{-4}\cos^{2}\beta,\nonumber \\
\varepsilon\psi & =\frac{1}{2}\left(\varepsilon-\frac{m}{2}\right)\left(\sin^{2}\beta-\frac{1}{3}\right)\Phi^{2}+\frac{m}{6}\Phi^{-3}\cos^{2}\beta,\label{eq:psi}
\end{align}
\begin{align}
\varepsilon\Phi^{-1}\frac{\partial\psi}{\partial\beta} & =X(\Phi)\cos\beta\sin\beta,\label{eq:dPsi_dbeta}\\
\text{where }X(\Phi) & =\left(\varepsilon-\frac{m}{2}\right)\Phi-\frac{m}{3}\Phi^{-4}.\label{eq:X}
\end{align}
(\ref{eq:psi}-\ref{eq:X}) specify completely (\ref{eq:map_TD15}).

\section{TD14 metric}

Rather than the mapping $(\lambda,\beta,\Phi)\mapsto\mathbf{r}$,
the equations of motion involve the metric $\text{d}l^{2}$. By design
$\text{d}l^{2}$ is, to $O(\varepsilon)$ accuracy, orthogonal, i.e.
$\text{d}l^{2}=\text{d}l_{TD}^{2}+O(\varepsilon^{2})$ with :
\begin{equation}
\text{d}l_{TD}^{2}=h_{\lambda}^{2}\text{d}\lambda^{2}+h_{\beta}^{2}\text{d}\beta^{2}+g^{-2}\text{d}\Phi^{2}\label{eq:dl2_TD14}
\end{equation}
where the scale factors $h_{\lambda},\,h_{\beta}\text{ and }h_{\Phi}=g^{-1}$
are obtained now. Using
\[
\mathbf{r}=\left(R_{E}(\Phi)-\sin^{2}\beta\Delta R(\Phi)\right)\mathbf{e}_{r}-X(\Phi)\cos\beta\sin\beta\mathbf{e}_{\beta}
\]
and
\[
\frac{\partial\mathbf{e}_{r}}{\partial\lambda}=\cos\beta\mathbf{e}_{\lambda},\quad\frac{\partial\mathbf{e}_{r}}{\partial\beta}=\mathbf{e}_{\beta},\quad\frac{\partial\mathbf{e}_{\beta}}{\partial\lambda}=-\sin\beta\mathbf{e}_{\lambda}
\]
one finds
\begin{align*}
\frac{\partial\mathbf{r}}{\partial\lambda}= & \left(R_{E}(\Phi)-\Delta R(\Phi)\sin^{2}\beta+X(\Phi)\sin^{2}\beta\right)\cos\beta\mathbf{e}_{\lambda}\\
\frac{\partial\mathbf{r}}{\partial\beta}= & -2\Delta R(\Phi)\sin\beta\cos\beta\mathbf{e}_{r}\\
 & +\left(R_{E}(\Phi)-\Delta R(\Phi)\sin^{2}\beta-X(\Phi)\cos2\beta\right)\mathbf{e}_{\beta}\\
\frac{\partial\mathbf{r}}{\partial\Phi}= & -\left(-\frac{\text{d}R_{E}}{\text{d}\Phi}+\frac{\text{d}\Delta R}{\text{d}\Phi}\sin^{2}\beta\right)\mathbf{e}_{r}\\
 & +\frac{\text{d}X}{\text{d}\Phi}\cos\beta\sin\beta\mathbf{e}_{\beta}
\end{align*}
Focusing first on the horizontal metric :
\begin{align}
h_{\beta}= & R_{E}(\Phi)-\Delta R(\Phi)\sin^{2}\beta-X(\Phi)\cos2\beta\label{eq:h_beta_Phi}\\
h_{\lambda}= & \left(R_{E}(\Phi)-\Delta R(\Phi)\sin^{2}\beta+X(\Phi)\sin^{2}\beta\right)\cos\beta.\label{eq:h_lambda_Phi}
\end{align}
Comparing (\ref{eq:h_beta_Phi}-\ref{eq:h_lambda_Phi}) to (\ref{eq:metric_ellipsoid_1}-\ref{eq:metric_ellipsoid_4})
and noting that $R_{E}^{-1}X=\Phi X+O(\varepsilon)$, one concludes
that the geoid $\Phi=cst$ is an ellipsoid of semi-major (resp. semi-minor)
axis $R_{E}(\Phi)$ (resp. $R_{E}(\Phi)-\Delta R(\Phi)$) and that
the coordinate $\beta$ is related to the conformal latitude $\phi_{c}^{\Phi}$
on that ellipsoid by :
\begin{equation}
\beta=\phi_{c}^{\Phi}+\Delta\phi\sin\phi_{c}^{\Phi}\cos\phi_{c}^{\Phi}+O(\varepsilon^{2}),\qquad\Delta\phi=\Phi X(\Phi)\label{eq:beta}
\end{equation}
to $O(\varepsilon)$ accuracy. Especially on the reference ellipsoid
$\Phi=1+(\varepsilon+m)/3,$ :
\begin{align}
\beta= & \phi+\left(\varepsilon-\frac{5m}{6}\right)\sin\phi\cos\phi+O(\varepsilon^{2}).\label{eq:def_beta}
\end{align}
where $\phi$ is defined, as in the main text, as the conformal latitude
on the reference ellipsoid. Focusing next on gravity :
\begin{align}
g^{-1} & =-\frac{\text{d}R_{E}}{\text{d}\Phi}+\frac{\text{d}\Delta R}{\text{d}\Phi}\sin^{2}\beta,\nonumber \\
g & =g_{E}(\Phi)+\Delta g(\Phi)\sin^{2}\beta+O(\varepsilon^{2}),\label{eq:gravity_TD14}\\
\text{where}\qquad g_{E}(\Phi) & =\Phi^{2}+\frac{1}{3}\left(\varepsilon-\frac{m}{2}\right)\Phi^{4}+2m\Phi^{-1},\label{eq:gE}\\
\Delta g(\Phi) & =-\left(\varepsilon-\frac{m}{2}\right)\Phi^{4}+2m\Phi^{-1}.\label{eq:Delta_g}
\end{align}
Especially, on the reference ellipsoid :
\begin{align}
g(\Phi_{a},\beta) & =1+m-\left(\frac{5}{2}m-\varepsilon\right)\cos^{2}\beta+O(\varepsilon^{2}).\label{eq:gravity_ref-1}
\end{align}

To $O(\varepsilon)$ accuracy, (\ref{eq:dl2_TD14},\ref{eq:h_beta_Phi},\ref{eq:h_lambda_Phi},\ref{eq:gravity_TD14})
with definitions (\ref{eq:RE},\ref{eq:DR},\ref{eq:X},\ref{eq:gE},\ref{eq:Delta_g})
specify the metric associated to the TD14 mapping, while (\ref{eq:def_beta})
relates latitude $\beta$ to the conformal latitude on the reference
ellipsoid. 

\section{Near-surface metric}

In order to simplify the metric near the reference ellipsoid, we can
finally change coordinates from $(\lambda,\beta,\Phi)$ to $(\lambda,\phi,\Phi)$,
i.e. use the conformal latitude on the reference ellipsoid as latitudinal
coordinate. In these coordinates, with $O(\varepsilon)$ accuracy
:

\begin{align*}
h_{\lambda}= & \left(R_{E}(\Phi)-\Delta R(\Phi)\sin^{2}\phi+\left(X(\Phi)-X_{a}\right)\sin^{2}\phi\right)\cos\phi\\
h_{\phi}= & R_{E}(\Phi)-\Delta R(\Phi)\sin^{2}\phi-(X(\Phi)-X_{a})\cos2\phi\\
g= & g_{E}(\Phi)-\Delta g(\Phi)\sin^{2}\phi
\end{align*}
where $X_{a}=X(\Phi_{a})$. Let us now consider near-surface region,
where $\xi=\Phi_{a}-\Phi=O(\varepsilon)$. Then 
\begin{align*}
X(\Phi)-X_{a}= & -\xi\,\frac{\text{d}X}{\text{d}\Phi}(\Phi_{a})+\dots=O(\varepsilon^{2})\\
\Delta R(\Phi)-\Delta R(\Phi_{a})= & -\xi\frac{\text{d}\Delta R}{\text{d}\Phi}(\Phi_{a})+\dots=O(\varepsilon^{2})\\
\Phi^{4}\frac{\text{d}\Delta R}{\text{d}\Phi}(\Phi)-\Phi_{a}^{4}\frac{\text{d}\Delta R}{\text{d}\Phi}(\Phi_{a})= & O(\varepsilon^{2})
\end{align*}
(note that $X,\,\Delta R=O(\varepsilon)$) so that :
\begin{align*}
h_{\lambda}= & \left(1-\xi\frac{\text{d}R_{E}}{\text{d}\Phi}-\varepsilon\sin^{2}\phi\right)\cos\phi+O(\varepsilon^{2})\\
h_{\phi}= & 1-\xi\frac{\text{d}R_{E}}{\text{d}\Phi}-\varepsilon\sin^{2}\phi+O(\varepsilon^{2})\\
g= & g(\Phi_{a},\phi)-\xi\frac{\text{d}g_{E}}{\text{d}\Phi}+O(\varepsilon^{2})
\end{align*}
Since $\xi=O(\varepsilon)$, it is sufficient to evaluate $\text{d}R_{E}/\text{d}\Phi$
and $\text{d}g_{E}/\text{d}\Phi$ to $O(1)$ accuracy, i.e. $-\text{d}R_{E}/\text{d}\Phi=\Phi_{a}^{-2}=1+O(\varepsilon)$,
$\text{d}g_{E}/\text{d}\Phi=2\Phi_{a}=2+O(\varepsilon)$, yielding
:
\begin{align*}
h_{\lambda}= & \left(1+\xi-\varepsilon\sin^{2}\phi\right)\cos\phi+O(\varepsilon^{2})\\
h_{\phi}= & 1+\xi-\varepsilon\sin^{2}\phi+O(\varepsilon^{2})\\
g= & 1-2\xi+m-\left(\frac{5}{2}m-\varepsilon\right)\cos^{2}\phi+O(\varepsilon^{2}).
\end{align*}

\end{document}